# Intense pulsed helium droplet beams

Mikhail N. Slipchenko[1], Susumu Kuma[2], Takamasa Momose[2], and Andrey F. Vilesov[1]

[1]Department of Chemistry, University of Southern California, Los Angeles, CA 90089, USA

[2]Permanent address: Division of Chemistry, Graduate School of Science, Kyoto University, Kyoto, 605-8502, Japan

Abstract

Pulsed (30 - 100 µs) nozzle beams have been used to generate helium droplets ($<N> = 10^4$-$10^5$). The dependence of the beam intensity and the mean droplet size on the source stagnation pressure and temperature are studied via mass spectroscopy and laser induced fluorescence of embedded phthalocyanine molecules. In comparison to a cw beam the pulsed source for the same pressure and temperature has a factor of 100 higher flux and the droplet sizes are an order of a magnitude larger.

Studies involving ultra cold, helium droplets have recently evolved into a dynamically developing field of research.[1-7] Helium droplets provide a uniquely gentle and very cold matrix for high resolution spectroscopy.[1,4,6] In this technique a single molecule or a well-defined number of molecules are embedded inside microscopic droplets of liquid $^4$He or $^3$He consisting of between $<N> = 10^3 – 10^5$ atoms.[2,8,9] The droplets are produced in a continuous (cw) cryogenic nozzle beam expansion.[2,8-11] Embedding is achieved by passing the droplets through a scattering chamber containing the species of interest.[2,12] Closed shell molecules are found to reside in the interior of the droplets.[1,2,4,13] The resonant absorption of light has been detected either by laser induced fluorescence or via the heat induced evaporation of several hundreds of weakly bound helium atoms which leads to a depletion of the droplet beam signal.[1,4,6] The special gentle nature of the matrix has been particularly impressively demonstrated in the infrared spectra of $SF_6$ [14-16] and $OCS$[17,18] and other molecules[1,4] by the resolution of sharp rotational lines with linewidths of about 150-300 MHz. From the intensities of the rotational transitions, temperatures between 0.38 K for pure $^4$He down to 0.15 K for the droplets with an excess of $^3$He have been obtained.[14-18] Other spectroscopic experiments indicate that $^4$He droplets are superfluid[19] and the ability for molecules to rotate freely in helium droplets appears to be a microscopic manifestation of superfluidity.[18] Effective cooling and very narrow linewidth of different species in helium promote the droplets as a useful experimental tool for the study of large organic molecules,[6,20] metal clusters,[21-25] and molecular complexes.[26-33] Simple chemical events in helium droplets such as the recombination of alkali atoms[21,22] and the reaction of Ba with $N_2O$[34] have also been documented. The fluid state of helium droplets is attractive, because atoms and molecules can move inside the droplet unhindered, and thus two or more species embedded independently will come together. This opens up the possibility to study ultra low temperature chemical processes in helium droplets, which may lead to new mechanisms and selectivity. Pulsed lasers are often used to induce photochemical reactions for the detailed study of state selective chemical reactions. On the other hand, the current cw sources of He droplets give relatively small droplet fluxes of about $10^{15} - 10^{16}$ droplets sterad$^{-1}$ s$^{-1}$.[35] Application of pulsed lasers to the cw beam obviously decreases the duty cycle of the experiment considerably. Therefore, the development of pulsed helium droplet technique is highly desirable. This can be mated with other pulsed experimental techniques such as: infrared depletion spectroscopy, laser-induced fluorescence (LIF), laser photolysis, laser ablation, and time-of-flight (TOF) mass spectroscopy, thus giving new interesting perspectives to the field.

The main reason pulsed helium droplet sources have not yet been introduced is the need for the pulsed valve to operate at low temperatures of $T_0 < 25$ K.[8,9,36,37] In particular, the sealing poppet of the valve must remain intact, which is difficult to achieve because most polymers crack at low temperatures. Operation of the solenoid type pulsed valves down to about $T = 100$ K have been previously shown.[38,39] Recently, Apkarian and coworkers[40] reported on the operation of a commercial solenoid type valve at temperature down to $T = 4$ K. This work encouraged us to replace the cw nozzle assembly in our helium droplet beam machine[20,27,33,34] with a pulsed nozzle source and test its performance. We used similar Series 99 (General Dynamics Valves Inc.) valve equipped with copper sealing gasket, which was operated by the commercial IOTA ONE (General Valves) driver. The shape of the nozzle channel was modified. The original 0.25 mm nominal diameter nozzle has a 0.5 mm diameter opening and a constriction in the outlet, see Fig. 1, insert (a). This configuration caused heating of the gas cooled during the initial stages of the expansion and a concomitant destruction of the droplets. As a result, the intensity of the droplet beam was small and it required a low operating temperature of about $T_0 = 8$ K. Therefore, we machined a ~90° conical opening which extends on the full length of the nozzle head of about 3 mm, see Fig. 1, insert (b). This modified valve gave more than a factor of 10 larger LIF intensity and higher operation temperatures (12-20 K).

A schematic of our experimental arrangement is shown in Fig. 1. In order to cool the stainless steel head of the valve, it is encased in a copper assembly, which in turn is attached to the cold head of a flowing liquid helium cryostat. In order to improve thermal contact, Apiezon grease is applied between the copper assembly and the stainless steel nozzle head. The thermal link is achieved by three 10 mm diameter copper rods, each 8 cm long. The temperature is actively stabilized via resistive heating of the cold head and is measured by two silicon diodes attached to the cold finger and the head of the valve, respectively. A large temperature difference of about 4-7 K between the cold finger and the head of the valve indicates an insufficient thermal link provided by the present construction. We anticipate that the largest improvement can be achieved by replacing the original General Valve stainless steel nozzle head with one made from copper, which has ~10 - $10^2$ larger heat conductivity.

We found that the valve starts opening typically at the driving pulse length of longer than $\Delta t$ = 150 μs, as monitored by the pressure rise in the nozzle chamber and appearance of the droplet signal in LIF- and mass- spectra. The maximal LIF signal corresponds to an incomplete valve opening at $\Delta t$ = 170 – 190 μs. This was found to vary somewhat upon the change of the poppet and related reassembling of the pulsed valve. The valve is opened completely at $\Delta t \approx$ 300 μs, as indicated by about 50 times increase in the gas throughput. No substantial increase of the peak droplet signal has been detected in this case. In part this is due to the distribution of the signal over the longer pulse. A longer pulse may also interfere with the gas reflected from the wall of the nozzle chamber.

The whole cryostat is mounted on an XYZ movable assembly. At 5 cm downstream from the nozzle the central part of the beam is separated by a 2 mm diameter, 2.5 cm long skimmer (Beam Dynamics). After a flight distance of 15 cm from the source, the droplets pass through a 2-cm-long heated (T = 450 $^0$C) scattering cell, which contains phthalocyanine. Gaseous Ar was admitted to the experimental chamber at pressures up to $10^{-5}$ mbar. The laser beam from a cw diode laser (TUI Laser, $\nu$ = 15089.92 cm$^{-1}$, $\delta\nu \approx$ 0.01 cm$^{-1}$, I = 0.6 mW), suitable to excite the origin of the $S_1 \leftarrow S_0$ transition of phthalocyanine in helium intersects the droplet beam 6 cm downstream from the pick-up cell. In other experiments, an excimer pumped pulsed dye laser was used in the same spectral range and laser-induced fluorescence (LIF) was collected by a photomultiplier (Hamamatsu RS 943-02) at right angles to both the droplet and laser beams. LIF signal was recorded by using a digital oscilloscope and a boxcar integrator. An additional monitor was provided by an on-axis quadrupole mass spectrometer equipped with an electron beam ionizer, which was positioned about 110 cm downstream from the droplet source.

The lower panel of Fig. 1 shows typical TOF dependencies of the LIF signal using the cw laser excitation (left) and mass spectrometer signal due to He$_2^+$ ions at m/e = 8 (right), obtained at $P_0$ = 40 bar and $T_0$ = 14 K. The strong LIF signal obtained at the laser excitation frequency corresponding to the transition of embedded phthalocyanine molecules indicates the presence of the droplets in the beam. The mass spectrometer signal consists of two peaks. The first is identified with direct ionization following electron impact. The second peak in the mass spectrum originates from the production of neutral metastable He$_2$* by electron impact, which remain attached to the droplets.[10,11] The droplets with attached He$_2$*, propagate with the beam velocity to the electron multiplier, where they create secondary ions. The time interval between LIF and the first ion peak permits the precise determination ($\pm$ 1 m/s) of the mean velocity of the droplet beam ($v_b$,= 394 m/s, at $T_0$ = 14 K). Upon decrease of the nozzle temperature, the peaks shift to longer delay times, which corresponds to a decrease of the beam velocity, being 330 m/s for $T_0$ = 9 K.

The peak widths of both the LIF (~32 μs FWHM) and ion (~46 μs FWHM) signals in Fig. 1 are much narrower than the used length of the driving pulse of $\Delta t$ = 180 μs. The pulse of the droplets has a spatial extent of about 1 cm at a distance of 20 cm from the source. The different widths of the LIF and ion pulses permit us to estimate the spread of the beam velocity to be $\Delta v$ = 4 m/s, which indicates a speed ratio of $v/\Delta v \approx$ 100. Similar speed ratios have been obtained in cw beams for He droplets formed via gas phase expansion.[10,11]

A comparison of the magnitudes of the ion signals from the pulsed and cw measurements indicates a factor of 100 higher peak flux of the droplets from the pulsed source. Similar estimates have been obtained from the LIF intensity measurements. When the source was operated at 50 Hz ($P_0$ = 20 bar, $T_0$ = 12.5 K) the steady state pressure rise in the nozzle chamber, equipped with 3000 L/s diffusion pump, was about $5 \times 10^{-5}$ mbar, which is ten times lower than for the cw source at similar $P_0$ and $T_0$. The heat load, caused by the dissipation of the kinetic energy of the moving piston, at $T_0$ = 10 K is estimated to be 5 mJ per pulse at $\Delta t$ = 180 μs. Pulse-to-pulse fluctuations of both LIF and ion signals of less than 5 % were observed, which are comparable or less than the output power fluctuations of the pulsed lasers. With proper adjustment, no leak through the valve equipped with the Kel-F poppet was detected (nozzle chamber base pressure 2 $\times 10^{-7}$ mbar) at any operating temperature (8-300 K) at stagnation pressure up to 40 bar (the highest pressure used). Moreover, because of the large nozzle diameter, it is not expected to be clogged by frozen impurities, as is often the case with the 5 μm cw nozzle. No deterioration of the source performance was noticed after several weeks of operation (~5 $10^7$ pulses) at $T_0$ < 20 K.

The poppet was finally changed after about $10^8$ pulses, when we noticed that the gas load in the nozzle chamber increased by about a factor of 5. Moreover about 2 K lower nozzle temperature was required to obtain similar signal as compared to the fresh poppet. The disassembled poppet had a sharp circular, about 0.1 mm deep, groove in the place of its sealing against the nozzle head.

Figure 2 shows the comparison of the LIF spectra of phthalocyanine molecules embedded in helium droplets measured several months before on the same machine equipped with a cw nozzle (a)[41] versus a pulsed nozzle (b). The spectrum measured with the pulsed nozzle has about a factor of $10^2$ higher S/N. At the energy of the pulsed laser of 10 µJ, the band origin of the transition is power saturated. The smooth feature on the blue side of the origin corresponds to the excitation of the phonon wing.[19,42] The spectrum corresponds to the $S_1 \leftarrow S_0$ transition, which have previously been studied in seeded molecular beams.[43] An intense band at $\lambda = 662.7$ nm is assigned as a band origin of the transition. This assignment is corroborated by the fact that no substantial features to the red have been observed. In comparison some features at the red from the most intense band have been observed in the free nozzle beam experiments,[43] which have been thought to be due to the shifted band origin, hot bands, van der Waals clusters or due to impurities. At the droplet temperature of 0.38 K, the hot bands are expected to be completely suppressed. The line assignments to complexes of a chromophore molecule with some other particles may be easily tested by controlled picking up of the corresponding species. Indeed, the very weak feature at $\lambda = 663.9$ nm is assigned to complexes of phthalocyanine with $N_2$ molecules, which have been picked up by the droplet from the residual gas. The superior quality of the LIF spectra obtained with the pulsed nozzle permits easy interrogation of diverse van der Waals complexes. The study of the complexes of phthalocyanine with simple species as $H_2$, $N_2$, Ar and $H_2O$ will be reported separately.[44] Sharp lines on the blue side of the origin are due to vibronic transi-tions of the phthalocyanine molecule. These transitions have much lower Franck-Condon-Factors as compared to the band origin, and acquire similar intensity in the LIF spectrum due to the power saturation of the band origin. Some vibrational frequencies of the phthalocyanine molecule in the $S_1$ state are indicated in Fig. 2 in units of wavenumbers. Similar frequencies in the free molecule are shown in brackets. Small shifts of the vibrational frequencies in helium of about 1 cm$^{-1}$ indicate weak interaction of the molecule with the helium environment. Vibrational frequencies in helium and linewidths of vibronic bands will be discussed in a forthcoming publication.[44]

The average number size of the droplets <N> has been obtained from the dependence of the LIF intensity on the Ar pressure in the experimental chamber. The phthalocyanine pick-up efficiency was kept small so that less than 10% of the droplets in the whole studied range of <N> were doped. An additional pick-up of Ar atoms results in the formation of complexes of Ar with phthalocyanine molecules within the droplet, whose absorption frequency is red shifted by 16 cm$^{-1}$. Thus, the LIF intensity due to bare phthalocyanine molecule, $I_{LIF}$, decreases with Ar number density [Ar] as:[12]

$$I_{LIF}(Ar) = I_{LIF}(0) \exp(-\sigma \times [Ar] \times L \times F(v_{rel}, v_b)), \qquad (1)$$

$$\sigma = c \times <N>^{2/3},$$

where: $\sigma$ is the pick-up cross section, c is a proportionality constant, [Ar] is the number density of Ar atoms, L = 16 cm is the length of the Ar pick-up region before the droplet beam crosses the laser beam, $F(v_{rel}, v_b) \approx v_{rel}/v_b$ is a correction factor to account for the finite thermal velocity of the Ar atoms, $v_{rel}$ is an average relative collision velocity of the argon atoms and helium droplets and c is a proportionality constant. In order to account for the effects of the droplet size distribu-tion,[8,9] and radial density distribution within a droplet,[8] the mean droplet size obtained from eq. (1) was calibrated against measurements with a cw nozzle, giving the constant c in eq. (1). For $P_0$ = 20 bar and $T_0$ = 11 and 12 K the direct beam deflection measurements gave $<N>$ = $1.8 \times 10^4$ and $1.4 \times 10^4$ atoms, respectively.[8] The decrease of $\sigma$ upon pick up of the phthalocyanine molecule was ignored in the pulsed experiments due to the large size of the droplets. In the cw calibration experiments Ar atoms were picked up prior to the phthalocyanine molecules.

Figure 3a shows the temperature dependencies of the average droplet size at different indicated values of $P_0$. The results for the cw 5μm nozzle from Refs.[8,9] are shown for comparison. Figure 3b shows similar dependencies of the droplet´s flux, $\Phi(P_0, T_0)$, which was estimated from the peak value of the $I_{LIF}$. In these measurements the frequency of the diode laser was adjusted to yield the maximum intensity of the band origin for each experimental condition. The shape of the spectrum of the band origin depends weakly on the droplet size in the relevant range of $2 \times 10^4 - 10^5$ atoms.[41] Thus relation

$$I_{LIF} \propto \Phi(P_0, T_0) \times \sigma / v_b, \qquad (2)$$

approximately holds. The cross section $\sigma$ (see eq. (1)) takes into account the dependence of the pick up probability on the droplet size.

It is instructive to compare the measured droplet size dependencies with those obtained for the cw expansion through a thin walled (2 μm)[45] nozzle with 5 μm diameter.[8,9]

i. Droplets having $10^4$-$10^5$ atoms are produced in a pulsed expansion at temperatures which are 5 to 10 K higher than the cw nozzle at the same $P_0$.

ii.     Droplets obtained with the pulsed nozzle are about an order of a magnitude larger than those obtained with the cw nozzle at similar expansion parameters at $T_0 > 10$ K, $P_0 = 20$ bar.

iii.    Average droplet sizes <N> show strikingly different behaviour as a function of nozzle temperature.  Whereas in the case of cw expansion the size of the droplets increases gradually with decreasing temperature, the formation of droplets in the pulsed expansion sets in sharply. These droplets are already quite large (<N> =2-5× $10^4$) and there is no evidence for smaller droplets at higher temperature.  The $He_2^+$ ions appear in the mass spectra at slightly higher nozzle temperature (1-2 K) as compared with the LIF signal.  Beyond the threshold the droplets size first remains unchanged for high expansion pressures ($P_0 = 20$ and 40 bar), or even drops suddenly for $P_0 = 6$ bar.  With the following decrease in $T_0$ the mean droplet size increases, saturates however at low temperature, or even drops for the highest pressure $P_0 = 40$ bar.

iv.     There is no indication for a transition to a supercritical expansion[10,11] which corresponds to liquefaction of helium within the nozzle followed by fragmentation of the liquid on larger droplets.  This regime of the cw expansion is characterized by very rapid rise of the average droplet size from $10^4$ to $10^6$ atoms within a temperature interval of less than 0.5 K below some threshold temperature, *e.g.,* T ≈ 10 K for $P_0 = 20$ bar and T ≈ 12 K for $P_0 = 40$ bar, see Fig. 3. This shape of the temperature dependence makes it difficult to exploit this range of the droplet size in the cw expansion which may be desirable for some particular experiments.  For example, the narrowest lines and largest S/N in LIF spectra of large molecules are obtained for <N> ~$10^5$. This range is also ideally suited for hosting large molecules and growing large clusters having up to 100 entities inside helium droplets.

Both i and ii are in agreement with the larger diameter of the nozzle.  The existing scaling laws[46] predict approximately linear increase of the scaling parameter with the nozzle size, which is in qualitative agreement with the present observations. However, it is not clear whether these scalings can be applied to the poorly defined expansion geometry of the pulsed nozzle. The existence of the sharp onset of formation of large droplets may indicate the threshold for nucleation in the expanding gas. Because there are insufficient nucleation centers at high temperature, they grow into large droplets in the following stages of the gas expansion. Note that similar threshold dependence has been observed for formation of the $^3He_n$ droplets in the case of cw expansion,[16,47] where the formation of the nucleation centers is hampered by instability of small clusters, having n < 29.[48] With further decrease of the nozzle temperature the number of nucleation centers increases rapidly, which causes the heating of the expanding gas and decrease of the mean droplet size. Droplet size increases with further decrease of the nozzle temperature, until the size and the number of the droplets become limited by the considerable amount of gas condensed into droplets. This scenario is in accord with the behaviour of the droplet flux, which rises sharply at the threshold, see Fig. 3b. The leveling out and drop of the droplet size and flux at higher expansion pressures may indicate the onset of scattering under these experimental conditions. The large difference of the droplet size dependencies on the expansion parameters between the pulsed and cw sources may be related to the larger dimension of the pulsed nozzle and concomitant longer residence time of the gas in the expansion region. Due to the incomplete opening of the nozzle the actual expansion zone is not circular but has a form of a ring, having a width of about 20-100 μm and diameter of about 0.5 mm. The length of this circular channel is estimated to be about 0.5 mm, *i.e.,* ~250 times longer than of the cw ones.

In summary, the first generation of pulsed beams of helium droplets is reported and the performance of the source is studied. In the future we are planning to upgrade this experiment with other pulsed experimental techniques, such as infrared depletion spectroscopy, laser photoly-

- 11 -

sis, laser ablation, and time-of-flight (TOF) mass spectroscopy, which will considerably increase the versatility of the technique. One of the future goals is to achieve the production of droplets having about $5 \times 10^3$ helium atoms, *i.e.,* the range, which gives the best S/N for depletion spectroscopy.[15] Another important continuation of this work is to explore the operation of the valve in the regime of full opening as well as using larger nozzle diameters. An additional increase of the beam intensity by a factor of 10 – 100 is easy to imagine.

**Acknowledgments**

This work has been supported by the University of Southern California set up funds. The authors are thankful to A. Apkarian who drew our attention to his experiments with cryogenic operation of the valve. We are also grateful to J. P. Toennies for the loan of the experimental equipment and careful reading of the manuscript.

# Figure Captions

**Figure 1.** Upper panel: schematic of the molecular beam apparatus showing characteristic distances. $T_1$ and $T_2$ show locations of the two Si thermoelements. Inserts a) and b) show the shape of the commercial and modified nozzle, respectively. Lower panel: (left) LIF signal versus delay relative to opening the nozzle; (right) mass spectrometer signal (m/e = 8) versus delay. Zero time corresponds to the front edge of the 180 μs driving electrical pulse of the valve. The helium droplets were produced by expansion at $P_0 = 40$ bar and $T_0 = 14$ K.

**Figure 2.** Comparison of LIF spectra of phthalocyanine molecules embedded in helium droplets measured (a) with a cw nozzle, and (b) with a pulsed nozzle. In both cases a pulsed dye laser (10 μJ, 70 Hz) was used and operating parameters were optimized to achieve the best possible LIF signal. Long wavelength pass filter of 715 nm was placed in front of the photomultiplier to reduce laser stray light. Scan speed: 0.02 nm/s, boxcar averaging over 30 laser pulses. The transition frequencies relative to the band origin are listed in units of wavenumbers, the numbers in brackets are the corresponding frequencies of the free molecule.[43]

Figure 3. a) Nozzle temperature dependencies of the mean helium droplet size for $P_0 = 6$, 20 and 40 bar for the pulsed nozzle. The results for the cw expansion through a 5 μm nozzle[8,9] for $P_0 = $ 20 and 40 bar are shown by thin continuous and dashed lines.
b) Similar dependencies of the droplet flux in arbitrary units, as obtained from eq. (2).

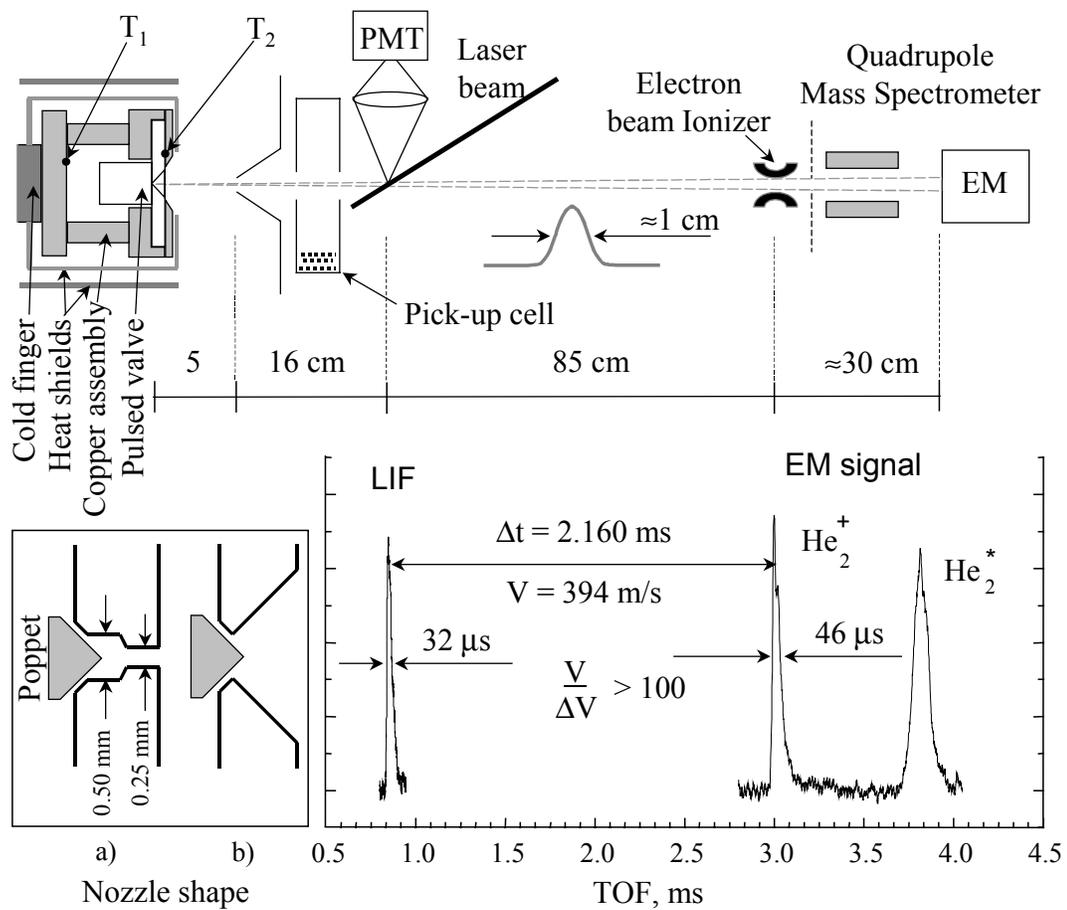

Figure 1.

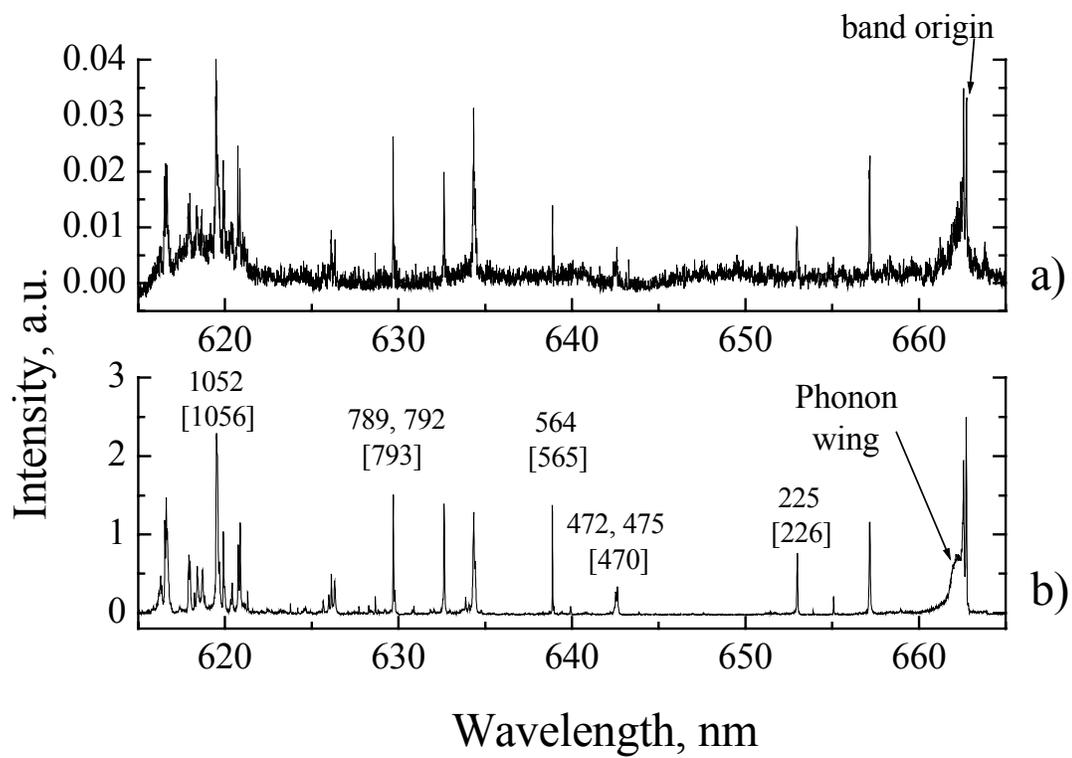

Figure 2.

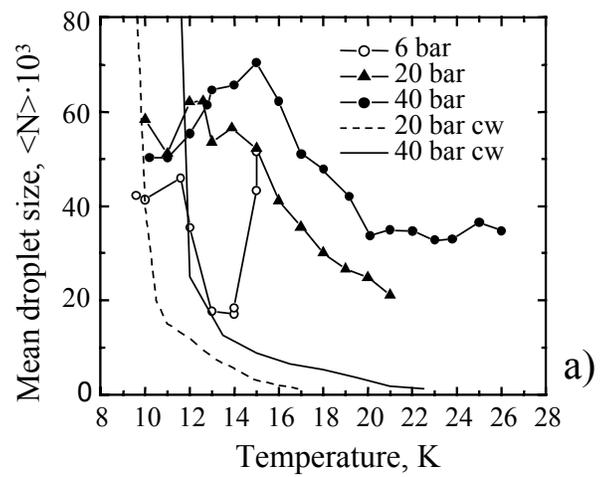

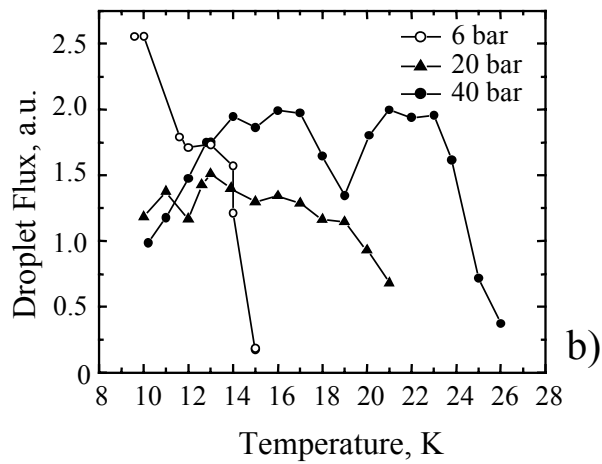

Figure 3.